\documentclass[12pt]{article}
\usepackage{fleqn}

\newcommand{\be}[1]{\begin{equation} \label{#1} }
\newcommand{\eq}{\end{equation}}
\newcommand{\bea}[1]{\begin{eqnarray} \label{#1} }
\newcommand{\eea}{\end{eqnarray}}
\newcommand{\eqrf}[1]{Eq.~(\ref{#1})}

\def\a{\alpha} \def\b{\beta}  \def\d{\delta}
 \def\f{\phi} \def\g{\gamma} 
 \def\j{\psi} \def\k{\kappa} \def\l{\lambda}
\def\m{\mu} \def\n{\nu} \def\o{\omega} \def\p{\pi} \def\q{\theta}
\def\r{\rho} \def\s{\sigma}  
 \def\z{\zeta}  
\def\G{\Gamma} \def\J{\Psi} \def\L{\Lambda} 
 \def\Q{\Theta}  \def\U{\Upsilon}

 \def\ce{\mathcal{E}} \def\cf{\mathcal{F}}
\def\cg{\mathcal{G}}  
  \def\cl{\mathcal{L}}
\def\cm{\mathcal{M}}  
  \def\car{\mathcal{R}}
\def\cs{\mathcal{S}} \def\ct{\mathcal{T}}

\def\pa{\partial}			
\def\bpa{\bar{\partial}}	 
\def\de{\nabla}				
\def\Iff{\Longleftrightarrow}		
\def\half{\frac{1}{2}}
\def\pr{^{\prime}}			
\def\prr{^{\prime\prime}}	
\def\tr{\mathrm{Tr}}			

\newcommand{\NPB}[1]{Nucl.\ Phys.\ \textbf{B#1}}
\newcommand{\PLB}[1]{Phys.\ Lett.\ \textbf{B#1}}

\newcommand{\xxx}[1]{\mbox{hep-th/\textbf{#1}}}

\def\MR{M.~Ro\v{c}ek {} }
\def\AGMR{A.~Giveon, \MR}

\makeatletter
\@addtoreset{equation}{section}
\@addtoreset{equation}{subsection}
\def\theequation{\ifnum\value{section}=0 \arabic{equation}\ignorespaces
\else \ifnum\value{subsection}=0 \thesection.\arabic{equation}\ignorespaces
\else \thesection.\arabic{subsection}.\arabic{equation}\ignorespaces
                       \fi
                 \fi}
\@addtoreset{table}{section}
\@addtoreset{table}{subsection}
\def\thetable{\ifnum\value{section}=0 \arabic{table}\ignorespaces
\else \ifnum\value{subsection}=0 \thesection.\arabic{table}\ignorespaces
\else \thesection.\arabic{subsection}.\arabic{table}\ignorespaces
                    \fi
              \fi}
\makeatother

\newcommand{\vier}[2]{{e^{#1}}_{#2}(x^\m)}
\def\myint{- \half\int d^{2}\s}
\newcommand{\suint}[1]{\myint #1_{+}#1_{-}}
\def\dil{\mathbf{\Phi}}
\def\metr{\mathbf{g}}
\newcommand{\riem}[1]{\mathbf{R}^{(-)}_{#1}}
\newcommand{\hriem}[1]{\hat{\mathbf{R}}^{(-)}_{#1}}
\newcommand{\conn}[3]{\mathbf{\omega}^{(#1)}_{#2,#3}}
\newcommand{\hconn}[3]{\hat{\mathbf{\omega}}^{(#1)}_{#2,#3}}
\def\hx{\hat{x}}
\def\hX{\hat{X}}
\def\jp{\psi_{+}}
\def\jm{\psi_{-}}
\def\Jp{\Psi_{+}}
\def\Jm{\Psi_{-}}
\newcommand{\hjp}[1]{\hat{\psi}_{+#1}}
\newcommand{\hjm}[1]{\hat{\psi}_{-#1}}
\def\hJp{\hat{\Psi}_{+}}
\def\hJm{\hat{\Psi}_{-}}
\def\pp{{+\!\!\!+}}
\def\dpp{\de_\pp}
\def\dmm{\de_{=}}
\def\one{\mathbf{1}}
\newcommand{\EB}[1]{\left(\mathbf{1} + \mathbf{b}\right)_{#1}}
\def\hE{\hat{E}}
\def\TT{\mathrm{T}}
\def\ppm{\pa_{\pm\pm}}

\begin{document}

\begin{titlepage}

\begin{flushleft}
ITP-SB-94-58 \\
hep-th/9411242 \\
November 30, 1994
\end{flushleft}

\begin{center}

\vskip3em
{\bf \huge
On Conformal Properties of \\ \vskip1ex
the Dualized Sigma-Models}

\vskip3em
{\Large\slshape Eugene Tyurin}

\vskip1em

{\slshape Department of Physics, SUNY at Stony Brook, \\ Stony Brook,
NY 11794-3800, USA \\ \vskip1ex E-mail: \hskip1ex
gene@insti.physics.sunysb.edu}

\end{center}

\vfill

\begin{abstract}
We have calculated the first-order $\b$-functions for a $\s$-model
( with dilaton) dualized with respect to an arbitrary Lie group
that acts without isotropy.  We find that non-abelian duality
preserves conformal invariance for semi-simple groups, but in
general there is an extra contribution to the $\b$-function
proportional to the trace of the structure constants, which cannot
be absorbed into an additional dilaton shift. Two particular
examples, a Bianchi V cosmological background and the
$\cg\otimes\cg$ WZW model, are discussed.
\end{abstract}

\vfill\hfill {\footnotesize Typeset by \LaTeXe}

\end{titlepage}

\section{Introduction}\label{intro}

The subject of target space duality (also called ``T-duality'')
has been important in the study of string models, because of, for
example, compactification issues and the inter-dependency of the
model dynamics on different scales and for different
manifolds. But historically, duality was performed with respect to
an abelian isometry group.

During the last couple of years, non-abelian duality received
significant attention. While formally rather similar to the
abelian case, it poses several problems that have been
successfully solved for abelian Lie groups, but become much more
difficult for non-abelian ones. The inverse transformation, the
transformation group, and the connection to the conformal
properties of the $\s$-models are among those problems.

This paper provides the lacking explicit proof \cite{T} that the
target space duality spoils the conformal invariance only if the
isometry group of the original conformal background has structure
constants that are not traceless.

The outline of the paper is as follows. In section \ref{NAD} we
briefly summarize the formulae relevant to the subject of
non-abelian duality transformations. For extensive reviews of both
abelian and non-abelian duality, see \cite{intro,T,big}. The
method that has been used to calculate the 1-loop $\b$-functions
is explained in section \ref{beta} and is followed by the explicit
derivation valid for \textbf{arbitrary} duality groups (that act
without isotropy). Following this, in section \ref{examples} we
work out two examples of $\s$-models that underwent non-abelian
duality transformation. The summary and discussion are in the
section \ref{fina}.

\section{Non-Abelian Duality}\label{NAD}

Suppose that we are given the usual $\s$-model action, with a
background that can be decomposed as follows ($\ce=\mathbf{g}+
\mathbf{b}$):\footnote{ As explained in \cite{AGMR}, this is the
most general case without isotropy; this restriction is made
purely to simplify the computations, but there is no reason to
believe that our results are not general.}
\bea{action}
\cs_\s &=& \myint \biggl[ \:
e^m(x^\m)\ce_{mn}(x^i)\bar{e}^n(x^\m) +
e^m(x^\m)\ce_{mj}(x^i)\bpa x^j \\ \nonumber
& &\qquad +\:\pa x^i\ce_{in}(x^i)\bar{e}^n(x^\m) +
\pa x^i\ce_{ij}(x^i)\bpa x^j - \frac{\sqrt{\g}}{4}\mathbf{R}^{(2)}\dil
\: \biggr],
\eea
where $e^m = \vier{m}{\mu}\pa x^\m \mbox{ and } \bar{e}^m =
\vier{m}{\mu}\bpa x^\m$ represent the coordinate frames that
obey the Maurer-Cartan equation
\be{M-C}
\hskip1.2in d e^m + \half f^m_{pq}e^p\wedge e^q = 0
\eq
with $f^m_{pq}$ being the structure constants of some Lie group $\cg$.
Equivalently, the generators of $\cg$
\be{gener}
\left[ \ct_m, \ct_n \right] = f^{p}_{mn}\ct_p, \qquad
\ct_m = {\q^\m}_m(x^\m)\:\pa_\m, \qquad
{\q^\m}_m = {\left( {e^m}_\m\right)}^{-1},
\eq
act as the Killing vectors on the background fields
$\ce$ and $\dil$ \cite{intro,dqc}:
\be{kill}
\cl_\ct\mathbf{g}=0, \qquad \cl_\ct\mathbf{b}=d\varpi,
\qquad \cl_\ct\dil = 0,
\eq
where $\varpi$ is some 1-form.
When the action $\cs_\s$ is invariant without total derivative
terms,
the $\cg$-isometry  of \eqrf{action} can be gauged by introducing
covariant derivatives with the $\cg$-valued gauge fields
$v_{\pm\pm} = v_{\pm\pm}^m\ct_m$
\be{covar}
\pa \rightarrow \de = \pa + v_\pp \qquad \mbox{ \ and \ }
\qquad \bpa \rightarrow \bar\de =\bpa + v_= ;
\eq
otherwise one needs to follow the procedure of \cite{iso}.
Note that $e^m$ by itself is a pure gauge:
\be{pure}
\vier{m}{\m}\pa x^\m = \tr\left(\ct^m g^{-1}\pa g\right),
\eq
which means that one can pick the ``unitary'' gauge \cite{que}
\be{ugauge}
g=1 \Iff x^\m=0 \quad\mbox{(in suitable coordinates)}
\eq
 for the model (\ref{action}), and reobtain the original
$\cs_\s$ by adding the following extra term to the action:
\be{lagr}
\cs_\l = \myint \left[\: \l_m \cf^m \:\right],
\mbox{ where }
\cf^m = \bpa v_\pp^m - \pa v_=^\m - v^p_\pp f^m_{pq} v^q_=
\eq
The equations of motion $\d \cs_\l /\d\l_m = 0$ require
the gauge field strengths $\cf^m$ to vanish, thus constraining
$v^m_\pp$ to ${e^m}$ and $v^m_=$ to ${\bar{e}^m}$.

To perform a \textbf{duality transformation} on the
$\s$-model described by \eqrf{action} and \eqrf{lagr} means to
promote the $\l_m$'s that were introduced as mere Lagrange
multipliers to regular coordinates of the \textbf{dual} space-time
by using the equations of motion for $v^m_\pp$, $v^m_=$ to
integrate $v^m_{\pm\pm}$ out of the action functional
\be{act}
\cs_\s\left[ x^\m = 0, x^i, v^m_{\pm\pm} \right] + \cs_\l
\eq
This transformation gives rise to the \textbf{dual} action:
\bea{action-dual}
& &\hspace{-3.5em}\hat{\cs} = \myint \Biggl\{
\left( \pa \l_m - \pa x^i \ce_{im} \right)
\left[ \left( \ce_{qq\pr} + \l_p f^p_{qq\pr}\right)
^{-1}\right]^{mn}
\left( \bpa \l_n + \ce_{nj}\bpa x^j \right) \nonumber \\
& &\quad +\:\ce_{ij} \pa x^i \bpa x^j -
\frac{\sqrt{\g}}{4}\mathbf{R}^{(2)}
\biggl[ \dil + \ln \det\left(\ce_{mn} +
\l_p f^p_{mn}\right)\biggr] \Biggr\}
\eea
It can be identified as the $\s$-model action with
the dual coordinates $\hx^m = \l_m$, $\hx^i = x^i$ and
the dual dilaton field
$\hat{\dil} = \dil + \ln \det\left(\ce_{mn}(\hx^i) +
\hx^p f^p_{mn}\right)$

For the rest of this paper, all the quantities arising from the
dual $\s$-model (\ref{action-dual}) will be labeled by ``hats''.

\section{Conformal Invariance and Sigma-Models}\label{conf}

If one thinks of $\s$-models as strings propagating in the
potential of some background fields (a space-time manifold and
dilaton matter comprise a particularly popular choice),
then consistency requires the
$\s$-model to be conformally invariant not only on the classical,
but also on the quantum level (for a review see
e.g. \cite{big,strings}). The 1-loop $\b$-functions are known to be:
\bea{beta1}
\b^\ce_{\a\g} &=& \riem{\a\g} - \de_\g\de_\a\dil \\
\b^\dil &=& \frac{1}{3\a\pr}(c-c_{crit}) +
\left[ (\de\dil)^2 + 2\de^2\dil -\riem{} -
\frac{1}{6}\mathbf{H}^2\right], \label{beta2}
\eea
where $\mathbf{H}_{\a\b\g}$ is the torsion tensor and
$\de$ and $\riem{}$ are the covariant derivative
and scalar curvature for the connection $\mathbf{\o}^{(-)}$
defined by \eqrf{defcn}. Note that $\b^\ce$ and $\b^\dil$ are not
independent, since, by using the Bianchi identities, one can
derive that, provided $\b^\ce_{\a\g} = 0$,
\be{boom}
\de_\g\b^\dil = -2 \de^\a\b^\ce_{\a\g} =  0
\eq
which means that $\b^\dil$ is a $c$-number and can always be set
equal to zero, once $\b^\ce_{\a\g}$ is known to vanish
\cite{strings}. Thus one can concentrate one's efforts on
\eqrf{beta1}.

\section{Calculating the Beta-Functions}\label{beta}

Probably the easiest way to calculate the dual $\b$-functions is
to use the Buscher's method \cite{bush}: to consider non-abelian
duality transformation of the N=1 supersymmetric extension of the
$\s$-model given by \eqrf{action}. We intend to benefit from the
well-known fact that the component action of N=1 $\s$-models
consists (after the elimination of the auxiliary fields) of purely
bosonic part corresponding to \eqrf{action}, 2-fermion coupling to
the bosonic connection and 4-fermion coupling to the bosonic
Riemann tensor:
\bea{components}
\hspace{-2em}\cs_{N=1} &=&  \suint{D}\biggl[\:
\ce_{\a\b}(\f ^\a) D_+ \f ^\a D_- \f ^\b \:\biggr]\nonumber \\
& = & \myint\biggl[\:
- \ce_{\a\b}(x^\a) \pa x^\a \bpa x^\b +
\half\jp^\a \jp^\b \jm^\g \jm^\d \riem{\a\b\g\d} \\
& + &  \jp^\a\bigl(\metr_{\a\b}\bpa \jp^\b +
\jp^\b \bpa x^\g \conn{-}{\a}{\b\g} \bigr) +
\jm^\a \bigl(\metr_{\a\b}\pa \jm^\b +
\jm^\b\pa x^\g\conn{+}{\a}{\b\g}\bigr)\:\biggr],\nonumber
\eea
where the connections in the cartesian coordinates are defined as
\be{defcn}
\conn{\pm}{\a}{\b\g} =
\half \left( \mathbf{g}_{\a\b ,\g} +\mathbf{g}_{\a\g ,\b} -
\mathbf{g}_{\b\g ,\a} \right) \pm
\half \left( \mathbf{b}_{\a\b, \g} +\mathbf{b}_{\g\a ,\b} +
\mathbf{b}_{\b\g ,\a} \right)
\eq
The following steps are involved in the procedure:
\begin{enumerate}
\item Write down the component action for the total
extended N=1 action $\cs = \cs_{\s,gauged} + \cs_\L$.
\item Dualize $\cs$ at the component level, with the unitary gauge
(\eqrf{ugauge} and (\ref{unit}) ) fixed.
\item Identify the dual connections and curvatures in terms
of the original ones, using the equations of motion for the super
gauge field.
\end{enumerate}
These steps are equivalent to dualizing the superfield $\L$ as a
whole, which produces the background fields identical to the ones
of a bosonic model \eqrf{action-dual}. This is the reason for the
validity of the relations for the connections and curvatures that
we obtain on the step 3.

We define our notation for the N=1 super Yang-Mills theory in 2
dimensions as follows ($\pa\equiv\pa_\pp$, $\bpa\equiv\pa_=$):
\bea{SYM}
\left\{ D_+ , D_- \right\} = 0 \qquad &
\left\{ D_\pm , D_\pm \right\} = 2 \pa_{\pm\pm} &\\
\de_\pm = D_\pm + \G_\pm \qquad &
\G_\pm = \G^m_\pm \ct_m \quad \mbox{( see \eqrf{gener})} & \nonumber \\
\left\{ \de_+ , \de_- \right\} \equiv W \qquad &
\left\{ \de_\pm , \de_\pm \right\} = 2 \de_{\pm\pm} =
2 \left( \pa_{\pm\pm} + \G_{\pm\pm} \right) &
\eea
The covariant component fields are defined as:
\bea{comps}
\f^\a\bigr| = x^\a & \de_\pm\f^\a\bigr| =\psi_{\pm}^\a &
\left( \de_\pm\f^i\bigr| = D_\pm\f^i\bigr| \right)
\nonumber \\
\L_m\bigr| = \l_m & \de_\pm\L_m\bigr| = \hat{\j}_{\pm m} & \\
& \de_\pm W\bigr| = G_\pm & \G_{\pm\pm}\bigr| = v_{\pm\pm} \quad
\G_\pm\bigr| = \z_\pm \nonumber
\eea
and the rules of covariant differentiation are
\bea{nil}
\left( \de_{\mp\mp} x \right)^\a &=&
\pa_{\mp\mp} x^\a + \d^a_\m v_{\mp\mp}^m {\q^\m}_m \\
\left( \de_{\mp\mp}\j_\pm \right)^\a &=&
\pa_{\mp\mp}\j_\pm^\a + \d^\a_\m
v_{\mp\mp}^m {\q^\n}_m \left( \pa_\n {\q^\m}_n \right)
{e^n}_\r\j_\pm^\r \nonumber
\eea
After removing the auxiliary fields
$W\bigr|$, $\de_+\de_-\L_m\bigr|$ and $\de_+\de_-\f^\a\bigr|$,
the N=1 extended action
\bea{1}
\cs &=& \cs_\s + \cs_\L \hfill\nonumber \\
& &\hspace{-4em} = \suint{\de}\left[ \:
\ce_{\a\b}\de_+\f^\a \de_-\f^\b \:\right]
\suint{\de}\left[ \L_m W^m \right]
\eea
becomes
\bea{2}
\cs_\s &=& \myint \Biggl[ \:
\half \jp^\a \jp^\b \jm^\g \jm^\d \riem{\a\b\g\d} \nonumber\\
&-& \ce_{\a\b}\dpp x^\a\dmm x^\b +
\mathbf{g}_{\a\b}\left( \jp^\a \dmm \jp^\b
+ \jm^\a \dpp \jm^\b \right) \nonumber \\
&+& \jp^\a\jp^\b\dmm x^\g\mathbf{\G}^{(-)}_{\a,\b\g} +
\jm^\a\jm^\b \dpp x^\g\mathbf{\G}^{(+)}_{\a,\b\g} \\
&-& \ce_{\a\b}\left( -\jp^\a G_-^m\q_m^\b + \nonumber
G_+^m\q_m^\a\jm^\b\right) \: \biggr] \\
\cs_\L &=&\myint \biggl[\: \hjp{m} G_-^m - \hjm{m} G_+^m \\
&+& \l_m \left( \bpa v^m_\pp -\pa v^m_= -
v^p_\pp f^m_{pq} v^q_= \right) + \nonumber
\l_m f^m_{pq} \left( G^p_+ \z^q_- - G^p_-\z^q_+ \right)\biggr],
\eea
after we have discarded the total derivatives
\be{totale}
\half\de_{\mp\mp}\left( \mathbf{b}_{\a\b}\j^\a_\pm \j^\b_\pm
\right) = \mathbf{b}_{\a\b}\j^\a_\pm\de_{\mp\mp}\j^\b_\pm +
\half\j^\a_\pm \j^\b_\pm\mathbf{b}_{\a\b,\g}\de_{\mp\mp}x^\g
\eq
The calculation will be significantly simplified in
orthonormal frames, which we always can choose to have the
following ``triangular'' form:
\bea{frames}
{E^M}_\m = {E^M}_m (x^i) \vier{m}{\m} &
\qquad {E^M}_i = {E^M}_i (x^i) \\
{E^I}_\m = 0 & \qquad {E^I}_i = {E^I}_i (x^i) \nonumber
\eea
Due to the above choice, the block form of $\Q = E^{-1}$ is:
\bea{tetas}
& {\Q^\m}_M = \left( {E^M}_\m\right)^{-1} &
\qquad {\Q^i}_M = 0 \\
& {\Q^\m}_I = - {\Q^\m}_M {E^M}_i {\Q^i}_I & \qquad
{\Q^i}_I = \left( {E^I}_i\right)^{-1} \nonumber
\eea
After fixing the unitary gauge (see \eqrf{ugauge})
\be{unit}
\f^\m = 0 \Longrightarrow x^\m=0 \mbox{ \ and \ }
\z^m_\pm = {e^m}_\m \j_\pm^\m \equiv \j_\pm^m
\eq
$\cs=\cs_\s+\cs_\L$ takes the following form:
\bea{orth}
\cs_\s &=& \myint \biggl\{ \:
\half\Jp^A\Jp^B\Jm^C\Jm^D\riem{ABCD} \nonumber \\
&-& v^p_\pp {E^P}_p \EB{PQ} {E^Q}_q v^q_= -
v^p_\pp {E^P}_p \EB{PB} {E^B}_j \bpa x^j \nonumber \\
&-& \pa x^i {E^A}_i \EB{AQ} {E^Q}_q v^q_= -
\pa x^i {E^A}_i \EB{AB} {E^B}_j \bpa x^j \nonumber \\
&+& \Jp^A\bpa\Jp^A + \Jp^A\Jp^B \left( \:
{E^C}_j \bpa x^j \conn{-}{A}{BC} + {E^Q}_q v^q_= \conn{-}{A}{BQ}
\: \right) \\
&+& \Jm^A\pa\Jm^A + \Jm^A\Jm^B \left( \:
{E^C}_i \pa x^i \conn{+}{A}{BC} + {E^P}_p v^p_\pp \conn{+}{A}{BP}
\: \right) \nonumber \\
&+& G^m_+ {E^M}_m \EB{MB} \Jm^B + G^m_- \Jp^A \EB{AM} {E^M}_m
\: \biggr\} \nonumber \\
\cs_\L &=& \myint \biggl[\: \hjp{m} G_-^m - \hjm{m} G_+^m \\
&+& \l_m \left( \bpa v^m_\pp -\pa v^m_= -
v^p_\pp f^m_{pq} v^q_= \right) + \nonumber
\l_m f^m_{pq} \left( G^p_+ \j^q_- - G^p_-\j^q_+ \right)\biggr],
\eea
For our future convenience, we define these matrices:
\bea{L}
\cl_{PQ} &\equiv& - {\Q^p}_P \left[\l_m f^m_{pq}\right]
{\Q^q}_Q \nonumber \\
\cm^{PQ} &\equiv& \left[\left(\EB{MM\pr} -
\cl_{MM\pr}\right)^{-1}\right]^{PQ} \\
\TT^{PQ} &\equiv& \left[ \mathbf{1} - 2\cm \right]^{PQ}
\quad \mbox{, note that }  \nonumber
\left( \TT \right)^{\mathrm{T}} = \left( \TT \right)^{-1}
\eea
The \textbf{dual} orthonormal frames are equal to:
\bea{dforms}
{\hE^P}{}_p = {\Q^p}_Q \cm^{QP} &
{\hE^P}{}_i = -\left({E^I}_i\mathbf{b}_{IQ} +
{E^{Q\pr}}_i\cl_{Q\pr Q}\right) \cm^{QP } \nonumber \\
{\hE^I}{}_p = 0 & {\hE^I}{}_i = {E^I}_i
\eea
with
\bea{11}
(\ppm\hX)^P =& {\hE^P}_p\ppm\l_p + {\hE^P}_i\ppm x^i
&\qquad (\ppm\hX)^I = (\ppm X)^I \\ 
\hat{\J}_\pm^P =& {\hE^P}_p\hat{\j}_{\pm p} + {\hE^P}_i\j_\pm^i
&\qquad \hat{\J}_\pm^I = \J^I_\pm 
\eea

The equations of motion for the gauge components
$v_{\pm\pm}$ and $G_\pm$ yield
\be{df}
\J_+^Q = \hat\J_+^Q \mbox{ \ and \ }
\J_-^Q = \TT^{QR} \hat{\J}_-^R
\eq
If we extend the definition of $\TT$ onto all indices as
following: $\TT^{IJ}=\one$ and $\TT^{IQ}=\TT^{PJ}=\mathbf{0}$, we
can compactly write
\be{dferm}
\J_+^A = \hat\J_+^A \mbox{ \ and \ }
\J_-^A = \TT^{AB} \hat{\J}_-^B
\eq
The contribution from the $v$-quadratic term equals
\bea{vv}
& \hspace{-2.5em} v^p_\pp {E^P}_p \cm^{-1}_{PQ} {E^Q}_q v^q_= =
\left[ (\pa\hX)^P - \pa x^i {E^P}_i +
\Jp^A \Jp^B \conn{-}{A}{BQ} \right] \times & \\
& \hspace{-2em}\cm^{QP} \left[ - (\bpa\hX)^R \cm^{-1}_{RP} -
\bpa x^j {E^R}_j \left( \one - \mathbf{b} + \cl \right)_{RP} +
\Jm^C\Jm^D\conn{+}{C}{DP} \right] & \nonumber
\eea
Finally, we can write down the dual component action:
\bea{aim}
& & \hspace{-3.9em}\hat\cs = \myint \biggl\{
-(\pa\hX)^P \cm^\TT_{PQ}(\bpa\hX)^Q
- (\pa\hX)^I\left(\one + \mathbf{b}\right)_{IJ}(\bpa\hX)^J
\nonumber \\
& & \hspace{-3em}-\:
 (\pa\hX)^P \cm_{PQ}{E^Q}_j{\hat\Q^j}_J(\bpa\hX)^J
+ (\pa\hX)^I{\hat\Q^i}_I {E^P}_i\cm^\TT_{PQ}(\bpa\hX)^Q \nonumber\\
& & \hspace{-3em}-\: (\pa\hX)^I{\hat\Q^i}_I\left(
{E^P}_i\cl_{PQ}{E^Q}_j + {E^P}_i\mathbf{b}_{PJ}{E^J}_j
 + {E^I}_i\mathbf{b}_{IQ}{E^Q}_j \right){\hat\Q^j}_J(\bpa\hX)^J
\nonumber \\
& & \hspace{-3em}+\:
 \hJp^A\bpa\hJp^A + \hJp^A\hJp^B\ \left[
\hconn{-}{A}{BJ}(\bpa\hX)^J +
\hconn{-}{A}{BP}\TT^{PQ}(\bpa\hX)^Q \right] \\
& & \hspace{-3em}+\:
 \TT^{AA\pr}\hJm^{A\pr}\pa\left(\TT^{AA\prr}\hJm^{A\prr}\right)
+ \TT^{AA\pr}\hJm^{A\pr}\TT^{BB\pr}\hJm^{B\pr}
\hconn{+}{A}{BC}(\pa\hX)^C \nonumber \\
& & \hspace{-3em}+\:
 \half\hJp^A\hJp^B\TT^{CC\pr}\hJm^{C\pr}\TT^{DD\pr}\hJm^{D\pr}
\left[ \riem{ABCD} + \conn{-}{A}{BP}2\cm^{PQ}\conn{+}{C}{DQ} \right]
\biggr\}. \nonumber
\eea
It gives us the sought relations:
\bea{result-}
\hconn{-}{A}{BC} &=& \conn{-}{A}{BC\pr}\TT^{C\pr C} \\
\hconn{+}{A}{BC} &=& \conn{+}{A\pr}{B\pr C}\TT^{A\pr A}\TT^{B\pr B}
+ \TT^{MA}{\hat\Q^c}_C \frac{\pa}{\pa\hx^c}\TT^{MB} \label{result+}\\
\hriem{ABCD} &=& \left[ \riem{ABC\pr D\pr} +
\conn{-}{A}{BP}2\cm^{PQ}\conn{+}{C\pr}{D\pr Q}\right]
\TT^{C\pr C}\TT^{D\pr D} \label{resultR}
\eea
The contribution to the dual $\b$-functions from the
dilaton shift $\hat\dil - \dil$ can be easily calculated if
one observes that
\be{shift}
\ln\det\left(\ce_{pq}+\l_m f^{m}_{pq}\right) =
\ln\det\left({E^M}_q{\hat\Q^p}_M\right)
\eq
Then, using the formula $\ln\det = \tr\ln$, we can
find that
\be{derivlndet}
\hat{\Q^a}_A \frac{\pa}{\pa \hx^a}
\ln\det\left(\ce_{pq}+\l_m f^{m}_{pq}\right) =
\left( \conn{-}{M}{AM} - \hconn{-}{M}{AM} \right) -
2 f^m_{mn}{\Q^n}_A
\eq
Also, it's easy to see that due to the relations \eqrf{kill} and
\eqrf{result-},
\be{dd}
\hat\de_D\hat\de_B\dil =
\left[\:\de_{D\pr}\de_B\dil\:\right] T^{D\pr D}
\eq
Now, using the fact that
\be{aaa}
\hspace{-2em}\hriem{BD} = \left(\riem{BD\pr} -
\riem{QBQD\pr}\right)\TT^{D\pr D}
+ \left( \conn{-}{I}{BP} - \hconn{-}{I}{BP} \right)
\hconn{-}{I}{PD} + \hriem{QBQD}
\eq
we arrive after some algebra to the central result of this work
\be{happy}
\hat\b_{BD} = \b_{BD\pr}\TT^{D\pr D} +
\hat\de_D\left[ 2 f^m_{mn}{\Q^n}_B(x^i)\right]
\eq
Note that for non-semi-simple Lie groups with $f^m_{mn}\neq 0$,
the right hand side of \eqrf{happy} can not
in general be compensated by adding an extra piece to the
dilaton field, unless for some particular case
\be{hmmmm}
{\Q^n}_B(\hx^i) = \hat\de_B\U(\hx^a)
\eq

\section{Applications}\label{examples}

In this section we demonstrate two particular cases of non-abelian
duality transformation that have been discussed in the literature.

First, the example of Bianchi V as a non-semi-simple subgroup of
the $SO(3,1)$ Lorentz group in 4 dimensions \cite{bian}. Second,
the duality of the vector-gauged $\cg\otimes\cg$ WZW model for
arbitrary semi-simple group \cite{AGMR}.

\subsection{Non-Semi-Simple Group}

In the paper \cite{bian}, the authors pointed out one particular
Bianchi-type non-semi-simple group, where conformal invariance of
the dualized $\s$-model could not be achieved by \emph{any}
correction to the dilaton. They discussed the following
initial background configuration:
\bea{venez}
& & x^\m = \left\{ x,y,z \right\} = \left\{ x^1,x^2,x^3 \right\}
\qquad x^i = \left\{ t \right\} = \left\{ x^4 \right\} \nonumber \\
& & f^2_{12} = f^3_{13} = 1 \quad \mbox{ the rest is zero } \\
& & \mathbf{g}_{\a\b} =
\mathrm{diag}\left( t^2, t^2 e^{-2x}, t^2 e^{-2x}, -1 \right)
\qquad \mathbf{b}_{\a\b} = \dil = 0 \nonumber
\eea
We have explicitly checked that an attempt to resolve the equation
\be{invain}
\hat\de_D \left( \hat\de_B \U - 2 f^m_{mn}{\Q^n}_B \right) = 0
\eq
leads to the same unsatisfiable condition $1/t^2=0$ that
have been reported in \cite{bian}.

\subsection{$\cg\mathbf{\otimes}\cg$ WZW Model}

The vector-gauged $\cg\mathbf{\otimes}\cg$ WZW Model for
semi-simple $\cg$ was considered in \cite{AGMR,que,iso} as an example of a
model without isotropy:
\be{qq}
\cs_{\cg\otimes\cg}[g,x] = \frac{k}{2\p}\int d^2\s\tr\Biggl[
\left( g^{-1}\pa g - x^{-1}\pa x \right)g^{-1}\bpa g\Biggr]
+ k \cs^{WZ}_\cg[x]
\eq
The dual action in this case is
\be{dqq}
\hspace{-1.5em}\hat\cs = \frac{k}{2\p}C^\car\!\!\int
d^2\s\left[\pa\l^m + (x^{-1}\pa x)^m\right]
(\one +\l_pf^p)^{-1}_{mn}\bpa\l^n
+ k \cs^{WZ}_\cg[x],
\eq
where in the group representation $\car$,
$\tr(\ct^\car_m\ct^\car_n)=C^\car\d_{mn}$ and the WZW action
\bea{wzw}
k \cs^{WZ}_\cg[x] &=& \frac{k}{2\p}\left[ \half\int d^2\s
\tr\left( x^{-1}\pa x\ x^{-1}\bpa x\right)+\G^{WZ}_\cg[x]\right]
\nonumber \\
&=& \frac{k}{4\p}C^\car\!\!\int d^2\s\biggl[
(x^{-1}\pa x)^m
\otimes (x^{-1}\bpa x)^m\\
&+&\left(\mathbf{b}^{WZ}_\cg\right)_{mn}
(x^{-1}\pa x)^m \wedge(x^{-1}\bpa x)^n\biggr] \nonumber
\eea
This calculation becomes relatively simple exercise in
differential geometry techniques \cite{eguchi} after one chooses
the frame
\bea{frr1}
E^M &=& {\k^M}_md\hx^m,\quad\mbox{ where }\k\equiv\left(\one +
\l_pf^p\right)^{-1} \mbox{ and } \hx^m\equiv\l_m \\
d E^M &=& \half{\left(\one + \k\right)^M}_n\ f^n_{PQ}\,
E^P\wedge E^Q \label{frr2}
\eea
Specifics of this concrete problem prompt us not to go for the
orthonormal frames. It's easy to see that $\k = \cm^\mathrm{T}$
(c.f. \eqrf{dforms}). The values of dual connections:
\renewcommand{\hconn}[3]{{\hat{\mathbf{\omega}}^{(#1)#2}{}}_{#3}}
\bea{ddd}
\hconn{-}{I}{JP} = {(\k -\one)^J}_{q}f^{q}_{IP} & &
\hconn{-}{I}{NP} = {(2\k -\one)^N}_{q}f^{q}_{IP} \nonumber\\
\hconn{-}{M}{JP} = -{\k^J}_{q}f^{q}_{MP} & &
\hconn{-}{M}{NP} = \left( {\k^M}_qf^q_{NP} -
2{\k^N}_qf^q_{MP}\right) \\
\hconn{-}{A}{BK} = 0 & &
\eea
agree with the values obtained by suitably modified
(c.f. \eqrf{frr1}) procedure, described above. The dual derivative
of the dilaton shift equals
\be{ooo}
\hat{Q}_{M}\ln\det\left(\one + \l_pf^p\right) = {\k^P}_Qf^Q_{PM}
\eq
One can check, using the Jacobi identities and the relations
\be{kkk}
\one -\k = \k\left(\l_mf^m\right) = \left(\l_mf^m\right)\k
\eq
that the dual background is, indeed, 1-loop conformally invariant.

\section{Conclusions and Discussions}\label{fina}

We have provided the proof that non-abelian duality respects the
conformal invariance of $\s$-models if the dualized isometry
corresponds to a semi-simple Lie group.  In the non-semi-simple
case, an anomaly \cite{gaume,intro} arises in the Jacobian of
duality transformation and violates the conformal invariance.
While formally our proof works only for the case without isotropy,
we strongly believe that our result should be valid in general. We
plan to address the isotropy case in the future.

Clearly, not all of the properties of abelian transformations can
be recovered in the much more complex environment of non-abelian
duality, which, unfortunately, is not sufficiently understood at
the present time.  Specifically, it is of great interest to
investigate the nature of the symmetries in the space of Conformal
Field Theories that are exhibited by the target-space duality.
Also, it seems to be useful to classify non-semi-simple groups
that due to some ``accident'' preserve the conformal invariance of
the original $\s$-model, since they might have some physical
significance \cite{bian}.

\section*{Acknowledgments}

It is a great pleasure to thank Martin Ro\v{c}ek for suggesting
this problem to me and for his guidance and advice.

\pagebreak

\end{document}